\documentclass[12 pt]{article}
\usepackage{amsmath,amssymb}
\usepackage{tikz}
\usepackage{natbib}
\usepackage{float}

\makeatletter

\floatstyle{ruled}
\newfloat{algorithm}{tbp}{loa}
\providecommand{\algorithmname}{Algorithm}
\floatname{algorithm}{\protect\algorithmname}

\makeatother

\newcommand{\dd}{\frac{\partial ^2}{\partial \psi^2}}

\newcommand{\bea}{\begin{eqnarray*}}
\newcommand{\eea}{\end{eqnarray*}}
\newcommand{\bean}{\begin{eqnarray}}
\newcommand{\eean}{\end{eqnarray}}
\newcommand{\bdm}{\begin{displaymath}}
\newcommand{\edm}{\end{displaymath}}

\def\boldfacefake #1{%
        \hbox{%
                \mathsurround=0pt
                \hbox to 0.25pt{$#1$\hss}%
                \hbox to 0.25pt{$#1$\hss}%
                \hbox {$#1$}%
        }%
}
{\hspace{-0.15 cm}
\begin{Sbox}
\begin{minipage}{12 cm}
 \vspace{0.15 cm} }%
{\vspace{0.15 cm} 
\end{minipage}
\end{Sbox}
\fbox{\TheSbox}}

\newcommand{\esl}{\end{slide}}
\newcommand{\bsl}{\begin{slide}}

\newcommand{\reals}{\mathbb{R}}

\newcommand{\iid}{\stackrel{\texttt{iid}}{\sim}}

\newcommand{\bx}{\mathbf{x}}
\newcommand{\bX}{\mathbf{X}}

\newcommand{\by}{\mathbf{y}}

\newcommand{\bY}{\mathbf{Y}}

\newcommand{\bz}{\mathbf{z}}

\newcommand{\bpsi}{\text{\mathversion{bold}$\psi$\mathversion{normal}}}

\newcommand{\bbeta}{\text{\mathversion{bold}$\beta$\mathversion{normal}}}

\newcommand{\bla}{\text{\mathversion{bold}$\lambda$\mathversion{normal}}}

\newcommand{\bte}{\text{\mathversion{bold}$\theta$\mathversion{normal}}}

\newcommand{\te}{\theta}
\newcommand{\la}{\lambda}

\newcommand{\bei}{\begin{itemize}}
\newcommand{\eni}{\end{itemize}}
\newcommand{\beq}{\begin{equation}}
\newcommand{\enq}{\end{equation}}

\newcommand {\bdot}{\hbox{\Huge .}}
\newcommand {\dotdot}{{\hbox{\Huge .}\kern-0.1667em\hbox{\Huge .}}}
\newcommand {\onedot}{1\kern-0.1667em\bdot}
\newcommand {\twodot}{2\kern-0.1667em\bdot}
\newcommand {\idot}{i\kern-0.1667em\bdot}
\newcommand {\jdot}{j\kern-0.1667em\bdot}
\newcommand {\mdot}{m\kern-0.1667em\bdot}
\newcommand {\dotj}{\kern-0.1667em\bdot\kern-0.1667em j}

\newcommand{\slopefrac}[2]{\leavevmode\kern.1em
\raise .5ex\hbox{\the\scriptfont0 #1}\kern-.1em
/\kern-.15em\lower .25ex\hbox{\the\scriptfont0 #2}}

\long\def\symbolfootnote[#1]#2{\begingroup%
\def\thefootnote{\fnsymbol{footnote}}\footnote[#1]{#2}\endgroup}

\begin{document}

\title{Approximate Integrated Likelihood via ABC methods }
\author{ Clara Grazian\thanks
{Corresponging Author: Department of Statistical Science.
Sapienza Universit\`a di Roma. 
Piazzale Aldo Moro, 5, 00185, Roma, Italy. CEREMADE Universit\'e Paris-Dauphine, Paris, France. CREST, Paris, France. e-mail: clara.grazian@ceremade.dauphine.fr
} \and Brunero Liseo\thanks{MEMOTEF, Sapienza Universit\`a di Roma, Viale del Castro Laurenziano 9, 00161, Roma, Italy.
e-mail: brunero.liseo@uniroma1.it. }
}
\date{\today} 
\maketitle

\symbolfootnote[0]{Acknowledgement: This work has been done with the contribution 
of grant ``Metodologie ABC (Approximate Bayesian Computation) 
per l'analisi di dati complessi", 2012 - n. C26A124RYM"
}

\begin{abstract}
We propose a novel use of a recent new computational tool for Bayesian inference, namely 
the Approximate Bayesian Computation (ABC)  methodology.
ABC  is a way to handle models for which the likelihood function may
be intractable or even unavailable and/or too costly to evaluate;
in particular, we consider the problem of eliminating the nuisance parameters from a 
complex statistical model in order to produce a likelihood function depending on the 
quantity of interest only.
Given a proper prior for the entire vector parameter, we propose to approximate the 
integrated likelihood by the ratio of kernel estimators of the marginal posterior and 
prior for the quantity of interest. We present several examples.
\end{abstract}

\noindent
{\em Key words: }{Monte\,Carlo methods, Nuisance parameters, Profile likelihood, 
Neyman-Scott
problem, Quantile estimation, Semi-parametric regression}.\\

\section{Introduction}
\label{1}

Given a statistical model with generic density $p(x\vert \bte)$, with
$\bte \in \Theta \subset \reals^d$, one is often interested in a low dimensional function
$\bpsi$ of the parameter vector $\bte$, say $\bpsi=\bpsi(\bte)\in \reals^k$, with $k<d$.
Modern parametric or semi-parametric statistical theories, at least the approaches based
on likelihood and Bayesian theories, aim at constructing a likelihood function which
depends on $\bpsi$ only. 
There is a huge literature on the problem of eliminating nuisance parameters, and we do
not even try to summarize it. Interested readers may refer to \citet{blw:99} and
\citet{enp:05} for a Bayesian perspective, and to the comprehensive books by \citet{ps:97} 
and
\citet{sev:00} or to \citet{lan:00} for a more classical point of view.
In a Bayesian framework the problem of eliminating the nuisance parameters is, at
least in principle, trivial. Let $\bla=\bla(\bte)$ the complementary parameter
transformation, such that $\bte=(\bpsi, \bla)$ and let 

\begin{equation}
\pi(\bte) = \pi(\bpsi, \bla) = \pi(\bpsi ) \pi(\bla \vert \bpsi ) 
\label{prior}
\end{equation}

\noindent
the prior distribution. Then, after assuming we observe a data set $\bx=(x_1, \dots,
x_n)$ from our working model, and computed the likelihood function $L(\bpsi, \bla; \bx)
\propto p(\bx; \bpsi, \bla)$, the marginal posterior distribution of $\bpsi$ is

\begin{equation}
 \begin{split}
\pi(\bpsi \vert \bx) = \frac{\int_\Lambda \pi(\bpsi, \bla) L(\bpsi, \bla; \bx)
d\bla}{\int_\Lambda\int_\Psi \pi(\bpsi, \bla) L(\bpsi, \bla; \bx) d\bla d\bpsi}
\\
\propto \pi(\bpsi) \int_\Lambda \pi(\bla \vert \bpsi) L(\bpsi, \bla;\bx) d\bla .
 \end{split}
\label{mp}
\end{equation}

The integral in the right-hand side of (\ref{mp}) is, by definition, the integrated
likelihood for the parameter of interest $\bpsi$, where ``integration'' is meant with respect to the conditional prior distribution $\pi(\bla \vert \bpsi)$; it will 
be denoted by $\tilde{L}(\bpsi; \bx)$.
The use of integrated likelihoods has become popular also among non Bayesian
statisticians; there are several examples in which its use is clearly superior, or
at least equivalent, even from a repeated sampling perspective, in reporting the actual
uncertainty associated to the estimates. See for example, \citet{sv3}, \citet{sv2} and 
\citet{sv1}.

However the explicit calculation of the above integral might not be so easy, especially 
when the dimension $d-k$ is large. Notice that the dimension $d$ may also include a 
possible latent structure which, from a strictly probabilistic perspective, is not 
different from a parameter vector.
In this paper we are interested to explore the use of approximate Bayesian computation
(ABC, henceforth) methods in producing an 
approximate integrated likelihood function, in situations where a closed form expression
of $\tilde{L}(\bpsi; \bx)$ is not available, or it is too costly even to evaluate the
``global" likelihood function $L(\bpsi, \bla; \bx)$, like, for example, in many genetic
applications or in the hidden (semi)-Markov literature. These are situations where MCMC
methods may not be satisfactory and completely reliable. 

Another class of problems where an integrated likelihood would be of primary interest is
that of 
semi-parametric problems, where 
the parameter of interest is a scalar - or a vector - quantity and the nuisance parameter
is represented by the nonparametric part of the model; in such cases the integration step
over the $\Lambda$ space would be infinite dimensional, and very often infeasible to be
solved in a closed form; we will discuss this issue in \S\, \ref{4}.

Approximate Bayesian computation has now become an essential tool for the analysis
of complex stochastic models when the likelihood function is unavailable. It can be
considered as a (class of) popular algorithms that achieves posterior simulation 
by avoiding the computation of the likelihood function (see \citet{bea:10}
for a recent survey).
A crucial condition for the use of ABC algorithms is that it must be relatively easy to 
generate new pseudo-observations from the working model, for a fixed value of the 
parameter vector. In its simplest form, the ABC algorithm is as follows (Algorithm 1 in \citet{mar:12})\\

\begin{algorithm}[!h]
\caption{Likelihood-free Rejection algorithm}

\label{ABC-rej}
\begin{flushleft}
\textbf{for} $i$ $=$  $1$ \textbf{to}  $N$  \textbf{do}

\textbf{repeat}

Generate $\te$ from the prior distribution $\pi(\te)$

Generate $z$ from the likelihood function $f (\cdot \mid \te )$

until $z = y$ (or some statistics $\eta$ is such that $\eta(z) \approx \eta(y)$) 
 
set $\te_i= \te$ 

\textbf{end for}
\end{flushleft}

\end{algorithm}

In this paper we will argue, through several examples of increasing complexity, how the
approximate integrated likelihood produced by ABC algorithms performs when compared with
the existing methods. We will also explore its use in particular examples where other
methods simply fail to produce a useful and easy-to-use likelihood function for the
parameter of interest.
The paper is organized as follows. In the next section we describe our proposal in
detail.
Section \ref{3} discusses some theoretical issues related to the precision of the ABC
approximation. Section \ref{4} compares the ABC integrated likelihood with other
available approaches in a series of examples. Section \ref{5} concludes with a final
discussion of pros and cons of the method.

\section{The proposed method}
\label{2}
The main goal of the paper is to obtain an approximation of the integrated likelihood
$\tilde{L}(\bpsi; \bx)$, for $\bpsi=\bpsi(\bte)\in \reals^k$.
From expression (\ref{mp})
it is easy to see that 
\beq
\label{core}
\tilde{L}(\bpsi; \bx) \propto \frac{\pi(\bpsi\vert \bx)}{\pi(\bpsi)},
\enq
that is the integrated likelihood function may be interpreted as the amount of
experimental evidence 
which transforms our prior knowledge into posterior knowledge about the parameter of
interest: from this perspective, we can interpret (\ref{core}) as the Bayesian definition 
of the integrated likelihood function.

Suppose that $\tilde{L}(\bpsi; \bx)$ is hard or impossible to obtain in a closed form.
For example the nuisance parameter $\bla$ might be infinite dimensional (see Example
4.4) or it may represent the non observable latent structure associated to the statistical model as in Hidden Markov or semi-Markov set-ups.

In these situations one can exploit the alternative expression (\ref{core}) of $\tilde
L(\bpsi; \bx)$. Of course, if $\tilde L(\bpsi; \bx)$ is not available, neither
$\pi(\bpsi\vert \bx)$ will be. However it is possible to obtain an approximate posterior
distribution $\tilde \pi(\bpsi\vert \bx)$, by using some standard ABC algorithm; in \S
\ref{3} we will discuss some issues related to the precision of this approximation; for
now, we describe the practical implementation of the method.
As in any ABC approach for the estimation of the posterior distribution, one has to
\bei
\item select a number of summary statistics $\eta_1(\bx), \dots, \eta_h(\bx)$;
\item select a distance $\rho(\cdot, \cdot)$ to measure the distance between ``true'' and proposed
data, or their summary statistics;
\item select a tolerance threshold $\varepsilon$

\item choose a (MC)MC algorithm which proposes values for the parameter vector $\bte$.
\eni
Once the posterior is approximated by  a size $M$ ABC posterior sample
$(\bte_1^\ast, \bte_2^\ast, \dots \bte^\ast_M)$, one can produce a non parametric kernel
based density approximation of the marginal posterior distribution of $\bpsi$, say 
$\tilde{\pi}^{ABC}(\bpsi\vert \bx)$. A similar operation can be done with the marginal
prior $\pi(\bpsi)$, by performing another - cheap - simulation from
${\pi}(\bpsi)$ to get another density approximation, say
$\tilde{\pi}(\bpsi)$. Notice that one is bound to use proper priors for all the involved 
parameters.

Then one can define the ABC integrated likelihood 
\beq
\label{defi}
\tilde{L}^{ABC}(\bpsi; \bx) \propto \frac{\tilde{\pi}^{ABC}(\bpsi\vert
\bx)}{\tilde{\pi}(\bpsi)}.
\enq

\section{The quality of approximation}
\label{3}
The gist of this note is to propose an approximate method for producing a likelihood 
function for a 
quantity of interest when the usual road of integrating with respect to the nuisance 
parameters 
cannot be followed.
There are two sources of error in (\ref{defi}). 
The first type of approximate error is 
introduced by the ABC approximation in the numerator so the level of
accuracy of (\ref{defi}) is of the same order of any ABC-type approximation.
We believe that the main difficulty with ABC methods is the choice of summary statistics.
However, while generic ABC methods have the goal of producing a ``global'' approximation to the posterior 
distribution, our particular use of the ABC approximation may suggest some alternative strategies for 
the choice of summary statistics. Classical statistical theory on the elimination of nuisance parameters can
be in fact of some guidance in the selection of summary statistics which are partially or 
conditionally 
sufficient for the parameter of interest. \citet{basu:77} represents an excellent reading on these topics.
In particular, his Definition 5 of ``Specific Sufficiency'' can be used in semi-parametric 
set-ups, like Example 4.4 below, where
the selected summary statistics are oriented towards the preservation of 
information about the parameter of interest. In our notation 
a statistic $T$ is specific sufficient for $\bpsi$ if, for each fixed value of the 
nuisance parameter $\bla$, $T$ 
is sufficient for the restricted statistical model in which $\bla$ is held fixed and 
known.

Another source of error in ABC is given by the tolerance threshold $\varepsilon$. As 
stressed in \citet{mar:12},
the choice of the tolerance level is mostly a matter of computational power: smaller 
$\varepsilon$'s are associated 
with higher computational costs and more precision. It is enough to reproduce the argument 
in \S\,1.2 of \citet{sis:11} to see that for $\varepsilon \to 0$, the error in 
(\ref{defi}), which is due to the tolerance, vanishes. 

Then, there is a balance between the fact that $\varepsilon$ has to be small and the fact that the simulation has to be practicable. It could be useful to choose $\varepsilon$ in a recursive way, by realizing a first simulation with a high tolerance level and then by choosing it in the left tail of the thresholds related to the accepted values. However, it is always recommended to compare different levels. 
  
The second main source of error is due to the kernel approximation step. A second order 
expansion for a Gaussian kernel estimator provides that

\begin{equation}
\mathbb{E}\left[\tilde{\pi}^{ABC}\left(\bpsi\vert\bx\right)\right]
=\pi\left(\bpsi\vert\bx\right)+\frac { 1 } { 2 } \dd 
\pi\left(\bpsi\vert\bx\right) h_x^2 k_2+\mathcal{O}\left(h_x^4\right)
\end{equation}

\noindent where $h_x$ is the bandwidth and $k_2=1$ in the case of Gaussian kernel.

A similar approximation holds for the prior distribution. 
Then, using general results on a first order approximation for the ratio of functions of 
random variables (\citet{ks1}, pag. 351), one has

\begin{equation}
\mathbb{E} \left[ \frac{ \tilde{\pi}^{ABC}\left( \bpsi\vert\bx\right) 
}{\tilde{\pi}\left(\bpsi\right)} \right]
=\frac{\pi\left(\bpsi\vert\bx\right)+\frac{1}{2} \dd \pi\left(\bpsi\vert\bx\right)
h_x^2+\mathcal{O}\left(h_x^4\right)}{\pi\left(\bpsi\right)+\frac{1}{2} \dd
\pi\left(\bpsi\right)h_\pi^2+\mathcal{O}\left(h_\pi^4\right)}
\end{equation}

\noindent where $h_x$ is the bandwidth chosen for the approximation of the posterior distribution and $h_\pi$ is the one chosen for the approximation of the prior. The prior distribution is often known in closed form or may be easily approximated with an higher accuracy than the posterior distribution.

The previous formula ensures that our estimator will be consistent provided that a sample size 
dependent bandwidth $h_n$, converging to $0$, is adopted.

It is a matter of calculation to show that the variance of the estimator is

\begin{eqnarray}
&& \phantom{             }     \mathbb{V}\left[ \frac{ \tilde{\pi}^{ABC}\left( 
\bpsi\vert\bx\right)}{\tilde{\pi}\left(\bpsi\right)} \right]  \\
&=&\left[ \frac{\pi\left(\bpsi\vert\bx\right)+C_\bx}{\pi\left(\bpsi\right)+C}\right]^2 
\nonumber
\\
&\times&
\left[ 
\frac{\frac{\pi\left(\bpsi\vert\bx\right)}{2nh\sqrt{\pi}}+\mathcal{O}\left(n^{-1}\right)} 
{\left[\pi\left(\bpsi\vert\bx\right)+C_\bx\right]^2}+ 
\frac{\frac{\pi\left(\bpsi\right)}{2nh\sqrt{\pi}}+\mathcal{O}\left(n^{-1}\right)} 
{\left[\pi\left(\bpsi\right)+C\right]^2}
\right] \nonumber 
\end{eqnarray}

\noindent where $$C_\bx=\frac{h_x^2}{2}\dd \pi \left( \bpsi\vert \bx\right) + 
\mathcal{O}\left( h_x^4\right)$$ 
and $$C=\frac{h_\pi^2}{2}\dd \pi \left( \bpsi\right) + 
\mathcal{O}\left( 
h_\pi^4\right)$$.

Again, using a bandwidth $h_n$, such that $h_n\to 0$, as $n\to \infty$, one can see that 
the first factor of the variance is asymptotically equal to the square of the true 
unknown value, while the second factor vanishes like $n^{-1}$.

In conclusion, the ABC approximation of the integrated likelihood function mainly depends on the ABC approximation and the kernel density estimate of the posterior distribution, whereas the prior distribution may be considered known, in general. \citet{blu:10} shows that the asymptotic variance of the kernel density estimator of the posterior distribution inversely depends on the number of simulations $n$ and on the kernel bandwidth, while the bias is proportional to the bandwidth. The mean squared error is minimized by 

\begin{equation}
h_n=\mathcal{O}\left(n^{-\frac{1}{d+5}} \right)
\end{equation}

\noindent where $d$ is the dimension of the summary statistics. Then the minimal MSE is 

\begin{equation}
MSE^*=\mathcal{O}\left(n^{-\frac{4}{d+5}}\right)
\end{equation}

\noindent which shows that the accuracy in the approximation decreases as the dimension of the summary statistics increases. This result may be used to define the number of simulations (and the burn-in) needed to reach the desired level of accuracy.

\section{Examples}
\label{4}
In this section we illustrate our proposal throughout several examples of increasing
complexity. The first one is a toy example and it is included only to show - in a very
simple situation - which are the crucial steps of the algorithm. \\
{\bf Example 4.1.}\,[Poisson means].
Suppose we observe a sample of size $n$ from $X\sim \mbox{Poi}(\te_1)$ and, independently
of it, another sample of size $n$  from $Y\sim \mbox{Poi}(\te_2)$. The parameter of
interest is $\psi=\te_1/\te_2$. This is considered a benchmark example in partial
likelihood literature since the conditional likelihood (see \citet{ks:70}), the profile
likelihood and the integrated likelihood obtained using the conditional reference prior 
\citet{blw:99} are all proportional to 
$$
\tilde{L}(\psi; \bx, \by) \propto \frac{\psi^{n\bar x}}{(1+\psi)^{n(\bar x+ \bar y)}},
$$
with the obvious meaning of the symbols above. Without loss of generality, set $\la=\theta_2$ as 
the nuisance parameter.

In this situation, the ABC approximation of the integrated likelihood is, in
some sense, not comparable with the ``correct'' integrated likelihood because
the latter is obtained through the use of an improper conditional reference
prior on $\la$ given $\psi$, and, as 
already stressed, it is not possible to use improper
priors in the ABC approach. A solution may be using a prior which mimics
the reference prior: we have taken 
$\te_1, \te_2 \iid  \mbox{Ga}\left(0.1,\,0.1\right)$. Notice that, 
in the economy of the method, only the prior on $\la$, not on $\psi$ is important.
The ABC
algorithm has been implemented to obtain approximations for the posterior
distributions of $\te_1$ and $\te_2$. The distance $\rho$ has been taken as the Euclidean 
distance, different
tolerance levels have been compared - $\varepsilon=\left(0.001,\,0.01,\,0.1,\,0.5\right)$ 
-
and the sample means of the two samples have been taken as summary
(sufficient) statistics. Samples of $1,000$ simulations have been
obtained to approximate the posterior distributions. The approximation
to the posterior distribution of $\psi$ is then simply obtained as
the ratio between the accepted values for $\te_1$ and $\theta_2$
via ABC. Given a sample from the prior distribution of $\psi$, the
approximation of its integrated likelihood is obtained through the
ratio between the kernel density estimates of both the prior and the
posterior distribution. 

\begin{figure}
\caption{\protect\includegraphics[width=9cm,height=8cm]{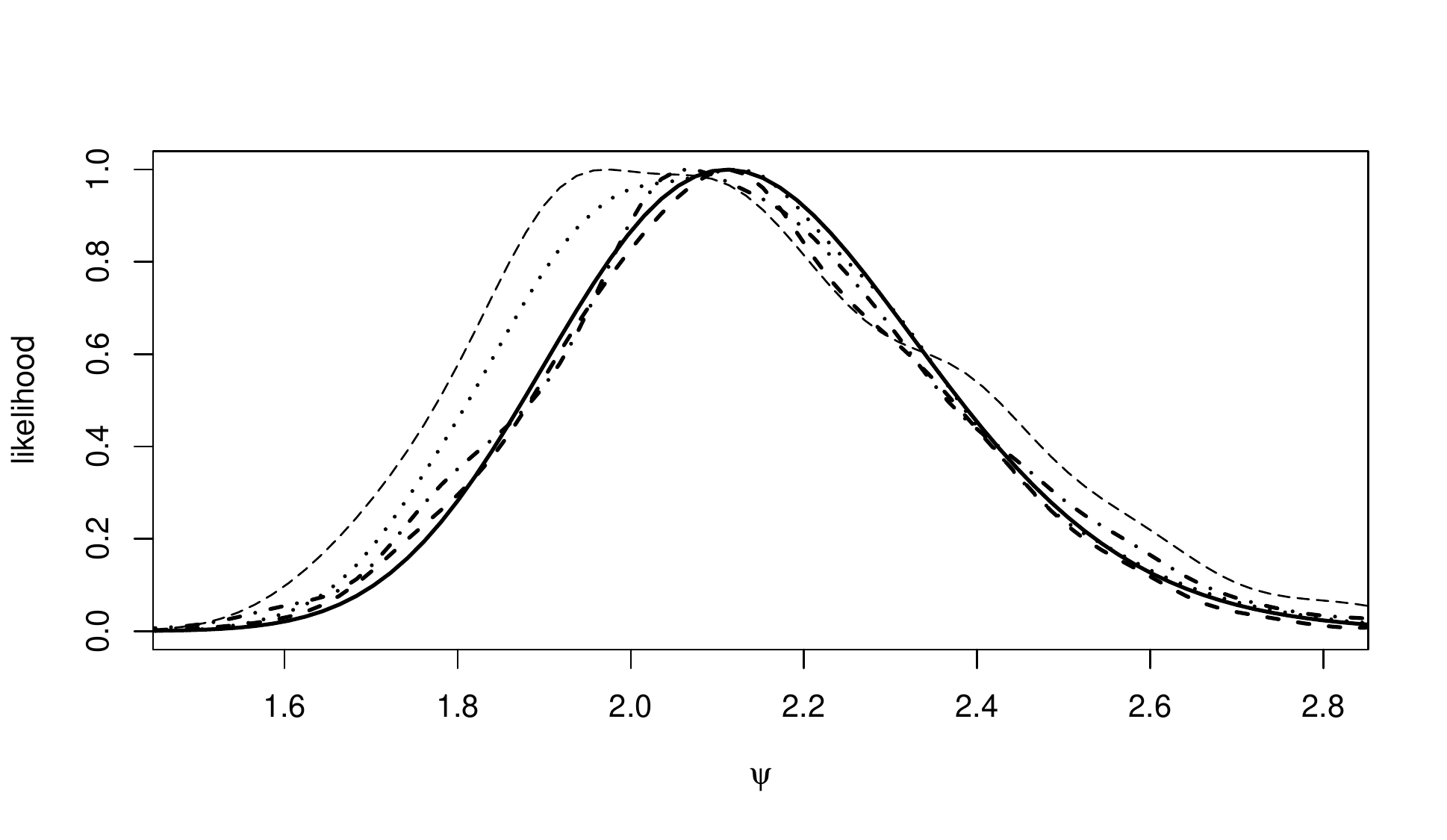}}
{\footnotesize The integrated likelihood of $\psi$ (solid line) and
its approximations for different tolerance levels: $\varepsilon=0.001$
(dashed), $\varepsilon=0.01$ (dotted), $\varepsilon=0.1$ (dotdashed) and
$\varepsilon=0.5$ (longdashed).}\label{fig confr toll gamma giusta}
\end{figure}

Figure \ref{fig confr toll gamma giusta} shows the approximations
with different choices of the tolerance level: the approximations are
close together and they are all close to the integrated likelihood;
the choice for the tolerance level does not seem to have a strong
influence; it is mostly a matter of computational power: the acceptance
rate is generally very low (often under 1\%), nevertheless it grows
with the tolerance level. As the threshold goes to zero, the approximation
is closer to the integrated likelihood, although the computational time
increases. 

Simulations have been repeated for different scenarios, by changing
the sample size and the number of simulations, however the results do
not seem to change in a significant way. In particular, as expected, the algorithm does 
not
depend on the  (induced) prior on $\psi$. 

{\bf Example 4.2.}\,[Neyman and Scott's class of problems].
This is a famous class of problems,
where the number of parameters increases with the sample
size (\citet{nesc:48}; \citet{lan:00}). Here we consider a specific example, already 
discussed in 
\citet{davi:03} and \citet{enp:05}, namely matched pairs of Bernoulli observations:
every subject is assigned to treatment or control group and the randomization
occurs separately within each pair, i.e. each data point in one data
set is related to one and only one data point in the other data set.
Let $Y_{ij}$'s be Bernoulli random variables, where $i=1,\dots,k$ represents
the stratum and $j=0,\,1$ indicates the observation within the pair.
The probability of success $p_{ij}$ follows a logit model:

\begin{equation}
\mathrm{logit}\, p_{ij}=\lambda_{i}+\psi_{j}
\end{equation}

\noindent For identifiability reasons, $\psi_{0}$ is set equal to $0$, while $\psi_{1}=\psi$
is considered constant across the $k$ strata; $\psi$ is
the parameter of interest. To formalize the problem, assume $\left(R_{i0},\,
R_{i1}\right)$ are $k$ independent matched pairs such that, for each $i$:

\begin{equation}
R_{i0}\sim\mathrm{Be}\left(\frac{e^{\lambda_{i}}}{1+e^{\lambda_{i}}}\right),
\quad
R_{i1}\sim\mathrm{Be}\left(\frac{e^{\lambda_{i}+\psi}}{1+e^{\lambda_{i}+\psi}}\right).
\end{equation}

\noindent The complete likelihood for
$\boldsymbol{\mathbf{\lambda}}=\left(\lambda_{1},\dots,\lambda_{k}\right)$
and $\psi$ is

\begin{equation}
L\left(\psi,\,\boldsymbol{\lambda}\right)=\frac{e^{\sum_{i=1}^{k}\lambda_{i}S_{i}+\psi
T}}{\prod_{i=1}^{k}\left(1+e^{\lambda_{i}}\right)\left(1+e^{\lambda_{i}+\psi}\right)}
\end{equation}

\noindent where $S_{i}=R_{i0}+R_{i1}$ for $i=1,\dots,k$ and $T=\sum_{i=1}^{k}R_{i1}$
is the number of successes among the cases. It is easy to show that
the conditional maximum likelihood estimate of $\lambda_{i}$ is infinite
when $S_{i}=0$ or $S_{i}=2$. The classical solution to this problem is to
eliminate the pairs where $S_{i}=0$ or $S_{i}=2$ from the analysis.
Nevertheless this is certainly a loss of information, because
the fact that a pair gives the same result under both treatments may suggest a ``not-so-big'' difference between groups.

It is easy to show that the 
conditional maximum 
likelihood estimator is 
$\left[\hat{\lambda}_{i,\psi}\mid\left(S_{i}=1\right)\right]=-{\psi}/{2}$; also, 
let $b$ be the number of pairs with $S_{i}=1$. The profile
likelihood of $\psi$ is 

\begin{equation}
\tilde{L}\left(\psi\mid S_{i}=1\right)=\frac{e^{\psi
T}}{\left(1+e^{\frac{\psi}{2}}\right)^{2b}}
\end{equation}

This likelihood function is not useful, since the maximum likelihood
estimate for $\psi$ is inconsistent (see \citet{davi:03}, Example
12.13): as $b$ increases, $\hat{\psi}\rightarrow2\psi$.
The modified version of the profile likelihood, proposed by \citet{bn:83} uses a 
multiplying factor:

\begin{equation}
M\left(\psi\right)=\left \vert 
J_{\lambda\lambda}\left(\psi,\,\hat{\lambda}_{\psi}\right)\right \vert^{-\frac{1}{2}}\left \vert 
\frac{\partial\hat{\lambda}}{\partial\hat{\lambda}_{\psi}}\right \vert=\frac{e^{\frac{b\psi}{4}}}{
\left(1+e^{\frac{\psi}{2}}\right)^{b}}
\end{equation}

\noindent where $J_{\lambda\lambda}\left(\psi,\,\hat{\lambda}_{\psi}\right)$
is the lower right corner of the observed Fisher information matrix. 

The conditional distribution of $T$ given $S_{1}=S_{2}=\dots=S_{b}=1$
is Binomial and depends on $\psi$ only. That is 
$T\mid [S_{1}=S_{2}=\dots=S_{b}=1,\,\psi] \sim
\mathrm{Bin}\left(b,\,\frac{e^{\psi}}{1+e^{\psi}}\right)$;
we can use it to get a conditional likelihood function:

\begin{equation}
L_{C}\left(\psi\right)\propto\tbinom{b}{T}\,\frac{e^{\psi
T}}{\left(1+e^{\psi}\right)^{b}}
\end{equation}

\noindent which leads to a consistent maximum conditional likelihood
estimator.

A Bayesian approach has the advantage that it does not need to discard
the pairs with $S_{i}=0$ or $2$. The likelihood contribution
for the $i$-th pair is simply

\begin{equation}
L\left(\psi,\,\lambda_{i}\right)=\frac{e^{\lambda_{i}S_{i}+\psi
R_{i1}}}{\left(1+e^{\lambda_{i}}\right)\left(1+e^{\lambda_{i}+\psi}\right)}.
\end{equation}

With a change of parametrization $\omega_{i}={e^{\lambda_{i}}}/{(1+e^{\lambda_{i}})}$
and using a (proper) Jeffreys' prior for $\omega_{i}\vert \psi$ (namely a $Beta\left(\frac{1}{2},\,\frac{1}{2}\right)$), 
the integrated likelihood is

\begin{equation}
L_{i}\left(\psi\right)=e^{\psi
R_{i1}}\int_{0}^{1}\frac{\omega_{i}^{S_{i}-\frac{1}{2}}\left(1-\omega_{i}\right)^{\frac{3}
{2}-S_{i}}}{1-\omega_{i}\left(1-e^{\psi}\right)}d\omega_{i}
\end{equation}

where the integral is one of the possible representation of the Hypergeometric or Gauss series,
as shown in \citet{abst:64} (formula 15.3.1, pag. 558). Therefore, the integrated 
likelihood is proportional to 

\begin{equation}
L_{i}\left(\psi\right)\propto \,_{2}F_{1}\left(1,\,
S_{i}+\frac{1}{2},\,3,\,1-e^{\psi}\right)\, e^{\psi R_{i1}}
\end{equation}

Define $L_{jl}\left(\psi\right)$ as the integrated likelihood function
associated with the $i$-th pair for which $\left(R_{i0},\, R_{i1}\right)=\left(j,\, l\right)$
and $n_{jl}$ the number of pairs for which $\left(R_{i0},\, R_{i1}\right)=\left(j,\,l\right)$, then the integrated likelihood function for $\psi$ is

\begin{equation}
L_{int}\left(\psi\right)\propto\prod_{j,l=0,1}L_{jl}\left(\psi\right)^{n_{jl}}.
\end{equation}

It is worthwhile to notice that this likelihood is not, in some sense,
comparable with profile and conditional likelihoods, because
it also considers the pairs discarded by non-Bayesian methods.

The ABC approach has been used with simulated data, with a sample
size $n$ equal to 30. Simulations were performed by setting $\psi=1$, a value which is
quite frequent in applications, when similar treatments are compared.
The values of 
$\boldsymbol{\lambda}=\left(\lambda_{1},\dots,\,\lambda_{n}\right)$ has
been generated by setting 
$\xi_i=\la_i/(1+\la_i)$ and drawing the $\xi'$s from a $ \mbox{U}(0,1)$ 
distribution.
Again, we have used the Euclidean distance between summary statistics and different tolerance levels
$\varepsilon=\left(0.001,\,0.01,\,0.1,\,0.5\right)$.
The summary statistics are the sample means for $\mathbf{R}_{0}$ and $\mathbf{R}_{1}$. We 
have also assumed a normal prior for $\psi$
with zero mean and standard deviation equal
to $10$. The proposed values for $\lambda_{i}$'s have been generated from 
a $Beta\left(\frac{1}{2},\,\frac{1}{2}\right)$ distribution for the above defined 
transformations $\xi_i$'s.

\begin{figure}
\caption{\protect\includegraphics[width=9cm,height=8cm]{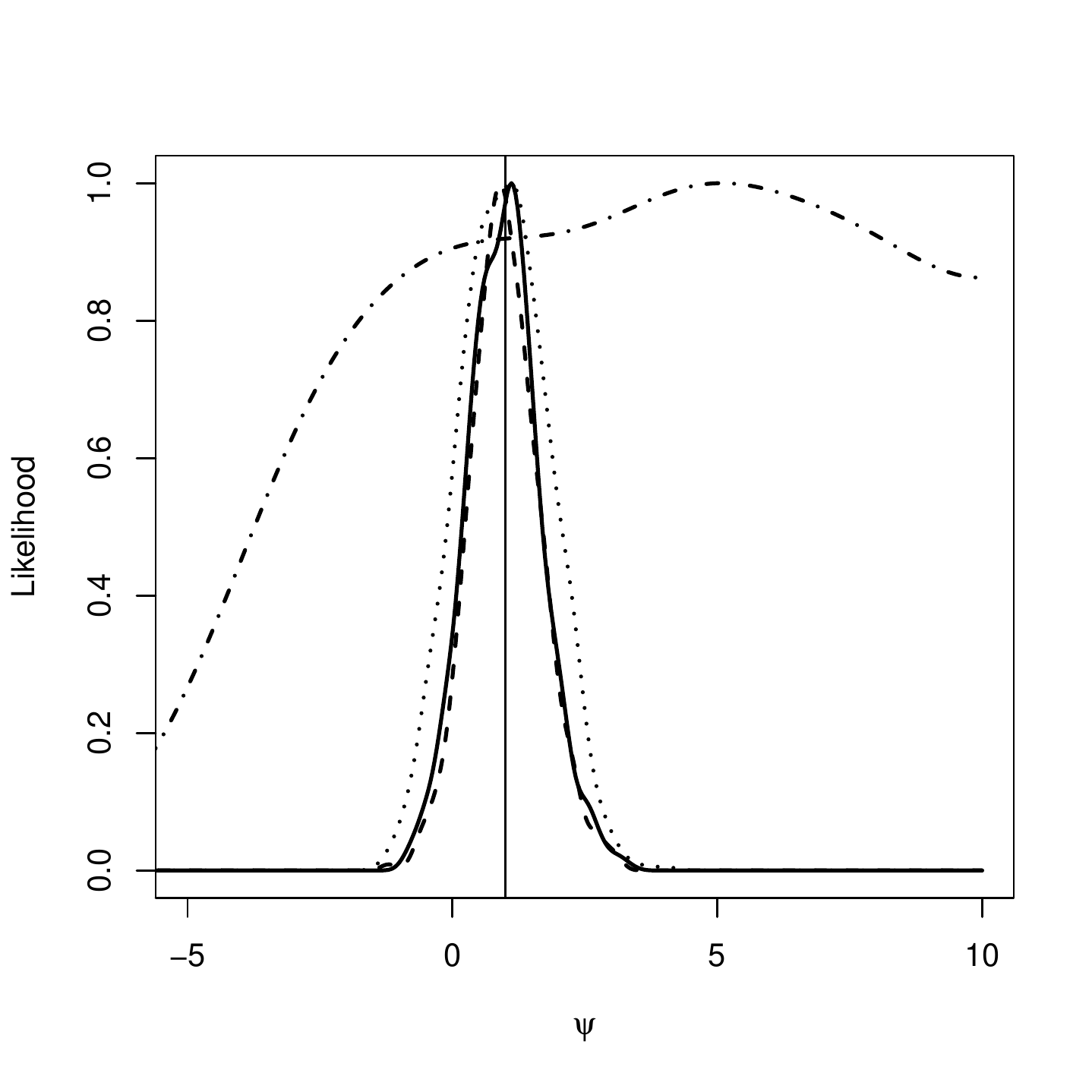}
\label{fig:matched pairs-confr toll-prior norm}}

{\footnotesize ABC approximations of the integrated likelihood for
$\psi$ with different tolerance levels: $\varepsilon=0.001$ (solid
line), $\varepsilon=0.01$ (dashed), $\varepsilon=0.1$ (dotted), $\varepsilon=0.5$
(dotdash).}
\end{figure}

With a sample from the posterior distribution of $\psi$ for each
tolerance level and a sample from its prior distribution, we have
obtained an approximation of the likelihood of $\psi$ via density
kernel estimation. The results are shown in Figure \ref{fig:matched pairs-confr toll-prior norm}:
the approximations are quite good for tolerance levels below 0.1; on 
the other hand, when
the threshold grows to 0.5 the approximate likelihood function is
very flat and multi-modal, i.e. too many proposed values, even very
different from the true value of $\psi$, are misleadingly accepted; for
example, a value of $\psi$ around $43$ has been accepted in one of our 
simulations. 

\begin{figure}
\caption{\protect\includegraphics[width=9cm,height=8cm]{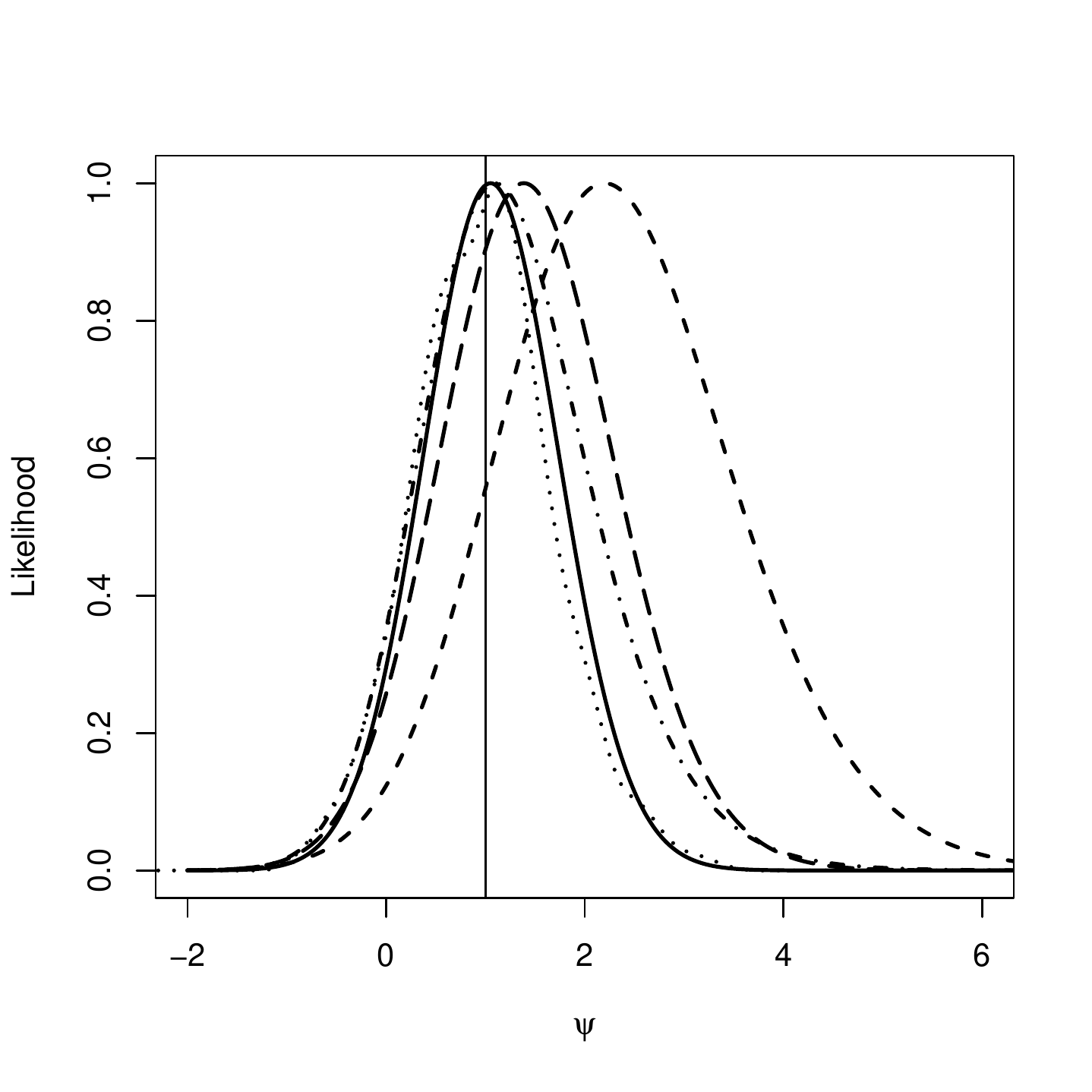}
\label{fig:matched pairs - confr lik prior normale}}

{\footnotesize Likelihood functions for $\psi$ based on different
solutions ($n=30$): the profile likelihood (dashed line), the conditional
likelihood (dotdash line), the modified profile likelihood (with the
correction of Barndorff-Nielsen, longdashed line), the integrated likelihood
(solid line, drawn by using the Laplace approximation of $_2F_1$ by \citet{butler:02}) and the ABC approximation ($n_{sim}=1000$, prior $\psi\sim
N\left(1,\,10\right)$
and tolerance level $\varepsilon=0.001$, dotted line).}
\end{figure}

Once again, a fair comparison between Bayesian and non-Bayesian approaches
is not strictly possible, nevertheless the various proposals are shown in Figure
\ref{fig:matched pairs - confr lik prior normale}: all the proposed solutions
are concentrated relatively close to the true value,
although the profile likelihood seems to be biased towards large values of $\psi$: 
this behaviour is also present, although to a minor extent in the modified profile and
the integrated likelihood solutions. The ABC approximation closely mimics 
the integrated likelihood, obtained via a saddle-point approximation of the Hypergeometric 
series \citet{butler:98}.

Similar conclusions are valid for different choices of the prior distribution,
different sample sizes, and different numbers of simulations. Just like in Example 4.1,
the acceptance rates are typically very low (always under $1\%$ for
tolerance levels under $0.01$ and about $5\%$ for a tolerance level
of $0.1$). Acceptance rates dramatically increase to about $60\%$, for $\varepsilon=0.5$; 
however in these cases, approximations get much worse.\\

\noindent
\textbf{Example 4.3} \,[Likelihood function for the quantiles of a $g$-and-$k$ 
distribution].
Quantile distributions are, in general, defined by the inverse of
their cumulative distribution function. They are characterized by
a great flexibility of shapes obtained by varying parameters values.
They may easily model kurtotic or skewed data with the great advantage
that they typically have a small number of parameters, unlike mixture models
which are usually adopted to describe this kind of data. 
An advantage of quantile distributions is that it is 
extremely easy to
simulate from them by means of a simple inversion. However, there are no free 
lunches, and the
above advantages are paid with the fact that 
their probability density functions (and therefore, the implied likelihood functions)
are often not available in a closed form expression. 

One of the most interesting examples of quantile class of distributions is the so-called 
$g$-and-$k$ distribution, described in \citet{hay:97}, whose quantile
function $Q$ is given by

\begin{equation}
\begin{split}
Q\left(u;\, A,\, B,\, g,\, k\right)=A+B\left[1+c\frac{1-\exp\left\{ -g\,
z\left(u\right)\right\} }{1+\exp\left\{ -g\, z\left(u\right)\right\} }\right]\cdot
\\
\left\{
1+z\left(u\right)^{2}\right\} ^{k}z\left(u\right)
\end{split}
\end{equation}

where $z\left(u\right)$ is the $u$-th quantile of the standard normal distribution;
parameters $A$, $B$, $g$ and $k$ represent location, scale, skewness
and kurtosis respectively; $c$ is an additional parameter which measures the
overall asymmetry and it is generally fixed at $0.8$, following \citet{ray2:02}.
 The class of Normal distributions is a proper subset of this class; it is obtained by
setting $g=k=0$.
Suppose we are interested in one or more quantiles using this model. There are no easy solution 
to the problem of constructing a partial likelihood for these quantiles.
The fact that the likelihood function is not available makes any classical approach practically 
impossible to implement. \citet{ray:02} propose
a numerical maximum likelihood approach; however they also explain that
very large sample sizes are necessary to obtain reliable estimates of the parameters.
On the other hand, even though this quantile distributions have no explicit likelihood, 
simulation from these models is easy, and an approximate Bayesian computation approach, 
also for producing an integrated likelihood of the parameters of interest,  
seems reasonable.

For this specific problem, two types of ABC algorithms have been compared: the former is
the usual ABC algorithm based on simulations from the prior distributions (with $10^{3}$
iterations); the latter is an ABC-MCMC algorithm ($10^{6}$ iterations, with a burn-in
of $10^{5}$ simulations). Two versions of ABC-MCMC have been used,
the former described in \citet{mar:12} (see Algorithm \ref{ABC-MCMC Robert})
and the latter described in \citet{all:09} (see Algorithm \ref{ABC-MCMC Mengersen}).
The main difference between these two versions of ABC-MCMC algorithm is that, in the first case,
there is no rejection step; at each iteration a value is accepted
(either the new proposed value or the value accepted in the previous iteration);
in the second case, instead, it is possible to discard the current value
and to propose a new one, so the chain always ``moves''. 

\begin{figure}
\caption{\protect\includegraphics[width=9cm,height=8cm]{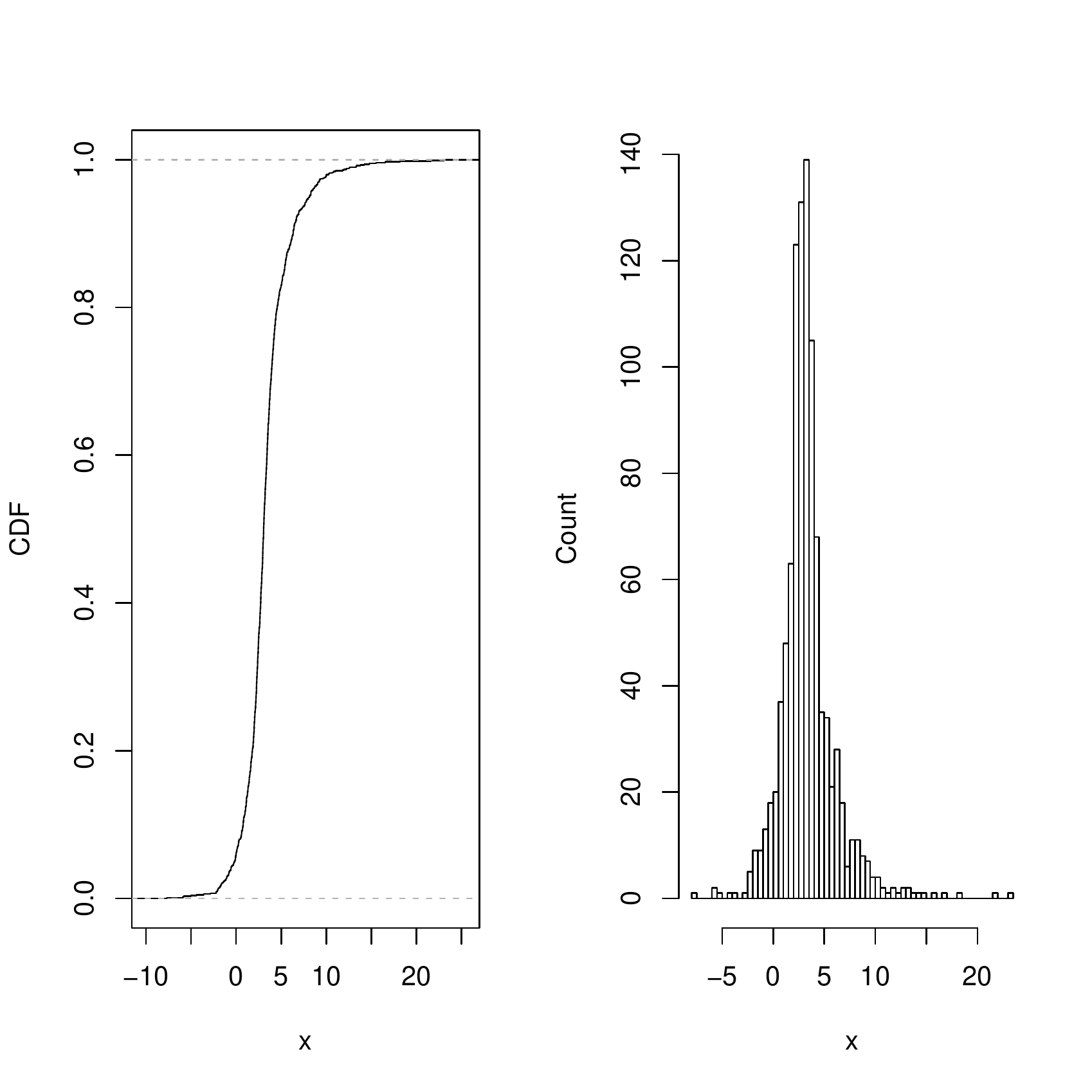}\label{fig:quantile-data}}

{\footnotesize Empirical cumulative distribution function (above) and histogram
(below) of the simulated data from a $g$-and-$k$ distribution (with
$A=3,\, B=1,\, g=2,\, k=0.5$). }
\end{figure}

Data have been simulated from a $g$-and-$k$ distribution with
parameters $A=3$, $B=1$, $g=2$ and $k=0.5$. As previously said,
$c$ is considered known and set equal to $0.8$. The sample size, 
has been set equal to $n=1000$. The empirical cumulative distribution function and 
the histogram
of the simulated data are shown in Figure \ref{fig:quantile-data}.

The transition kernel of the ABC-MCMC algorithm needs to be chosen having in mind two 
conflicting objectives: on one hand, full exploration of the parameter space, and, on the 
other hand, a reasonably high acceptance rate, which increases for proposals mostly 
concentrated where the posterior mass is present. 
As described in \citet{all:09}
uniform priors with bounds $\left(0,\,10\right)$ have been chosen for each
parameter and a random walk-normal kernel with variance $0.1$ has been used 
together with a large number of iterations ($10^{6}$) so that the parameter space 
is likely to be fully investigated. 
The vector of summary statistics consists of the sample mean, the standard
deviation, and the sample skewness and kurtosis indexes. The Euclidean distance 
has been used to compare summary statistics. 

The tolerance level $\varepsilon$ has been chosen in a recursive way:
first, a very large value has been selected, and a histogram of all
the distances has been drawn. A reasonable value has been taken from
the $5\%$ left tail of this histogram. Then, the chosen threshold 
has been compared with smaller values. In particular, a threshold
equal to $3$ corresponds to $3.9\%$ left tail. This has been compared
with tolerance levels equal to $2$ and $0.5$.

\begin{algorithm}[!h]
\caption{Likelihood-free MCMC sampling}

\label{ABC-MCMC Robert}
\begin{flushleft}
Initialization

A1) Generate $\theta'$ from the prior distribution $\pi\left(\cdot\right)$

A2) Generate a data set $\mathbf{z}'\sim f\left(\mathbf{z}\mid\theta'\right)$,
where $f$ is the model of the data

A3) If $\rho\left\{ \eta\left(\mathbf{y}\right),\,\eta\left(\mathbf{z}'\right)\right\}
\leq\varepsilon$,
set $\left(\theta^{(0)},\,\mathbf{z}^{(0)}\right)=\left(\theta',\,\mathbf{z}'\right)$,
otherwise return to A1)

MCMC-step

\textbf{for} $t=1,\,\dots,\, T$ \{

1) Generate $\theta^{prop}$ from the Markov kernel $q\left(\cdot\mid\theta^{(t-1)}\right)$

2) Generate $\mathbf{z}'$ from the model $f\left(\cdot\mid\theta^{prop}\right)$

3) Calculate
$h\left(\theta^{(t-1)},\theta^{prop}\right)=\min(1,\:\frac{\pi\left(\theta^{prop}\right)\,
q\left(\theta^{(t-1)}\mid\theta^{prop}\right)}{\pi\left(\theta^{(t-1)}\right)\,
q\left(\theta^{prop}\mid\theta^{(t-1)}\right)})$

4) \textbf{if} $\rho\left\{
\eta\left(\mathbf{y}\right),\,\eta\left(\mathbf{z}'\right)\right\} \leq\varepsilon$,
set
$\left(\theta^{(t)},\,\mathbf{z}^{(t)}\right)=\left(\theta^{prop},\,\mathbf{z}'\right)$
with probability $h$, 

\textbf{else}
$\left(\theta^{(t)},\,\mathbf{z}^{(t)}\right)=\left(\theta^{(t-1)},\,\mathbf{z}^{
\left(t-1\right)}\right)$

\}
\end{flushleft}
\end{algorithm}

\begin{algorithm}
\caption{Likelihood-free MCMC sampling}

\label{ABC-MCMC Mengersen}
\begin{flushleft}

Initialization

A1) Generate $\theta'$ from the prior distribution $\pi\left(\cdot\right)$

A2) Generate a data set $\mathbf{z}'\sim f\left(\mathbf{z}\mid\theta'\right)$,
where $f$ is the model of the data

A3) If $\rho\left\{ \eta\left(\mathbf{y}\right),\,\eta\left(\mathbf{z}'\right)\right\}
\leq\varepsilon$,
set $\left(\theta^{(0)},\,\mathbf{z}^{(0)}\right)=\left(\theta',\,\mathbf{z}'\right)$,
otherwise return to A1)

MCMC-step

\textbf{for} $t=1,\,\dots,\, T$ \{

1) Generate $\theta^{prop}$ from the Markov kernel $q\left(\cdot\mid\theta^{(t-1)}\right)$

2) Generate $\mathbf{z}'$ from the model $f\left(\cdot\mid\theta^{prop}\right)$

3)Calculate
$h\left(\theta^{(t-1)},\theta^{prop}\right)=\min(1,\:\frac{\pi\left(\theta^{prop}\right)\,
q\left(\theta^{(t-1)}\mid\theta^{prop}\right)}{\pi\left(\theta^{(t-1)}\right)\,
q\left(\theta^{prop}\mid\theta^{(t-1)}\right)})$

4) \textbf{ if} $\rho\left\{
\eta\left(\mathbf{y}\right),\,\eta\left(\mathbf{z}'\right)\right\} \leq\varepsilon$,
set
$\left(\theta^{(t)},\,\mathbf{z}^{(t)}\right)=\left(\theta^{prop},\,\mathbf{z}'\right)$
with probability $h$, 

\textbf{else} return to 1)

\}
\end{flushleft}
\end{algorithm}

The analysis of the approximate posterior distributions shows that
three out of four parameters ($A$, $B$ and $k$) are well identified,
while the posterior distribution of $g$ is rather flat. 
In general, as the tolerance level decreases,
results improve and posterior distributions tend to be more concentrated. 
Nevertheless, even using the lowest tolerance level the posterior
distribution of $g$ does not seem to concentrate around any value. 
This suggests that the algorithm needs an even smaller value of the threshold.
A simulation with tolerance level equal to $0.25$ has been then performed using 
Algorithm \ref{ABC-MCMC Robert}: the approximation of the posterior distribution
of $g$ is still not centred around its true value, even if there
is a mode around it; nevertheless the problem with this so low tolerance
level and this type of algorithm is that the acceptance rate of new
proposed values is very low and the chain does not move too much.
This tolerance level is also so low to make the application of the
other algorithms prohibitive in terms of computational time.

Our main goal of the analysis was to find an approximation of the
integrated likelihood function for a given quantile: in particular, we have
considered the percentiles of order $0.05$,
$0.10$, $0.25$ and $0.50$. Notice that, in the $g$-and-$k$
distribution model, the median is always equal to $A$.

\begin{figure}[h!]
\caption{\protect\includegraphics[width=9cm,height=8cm]{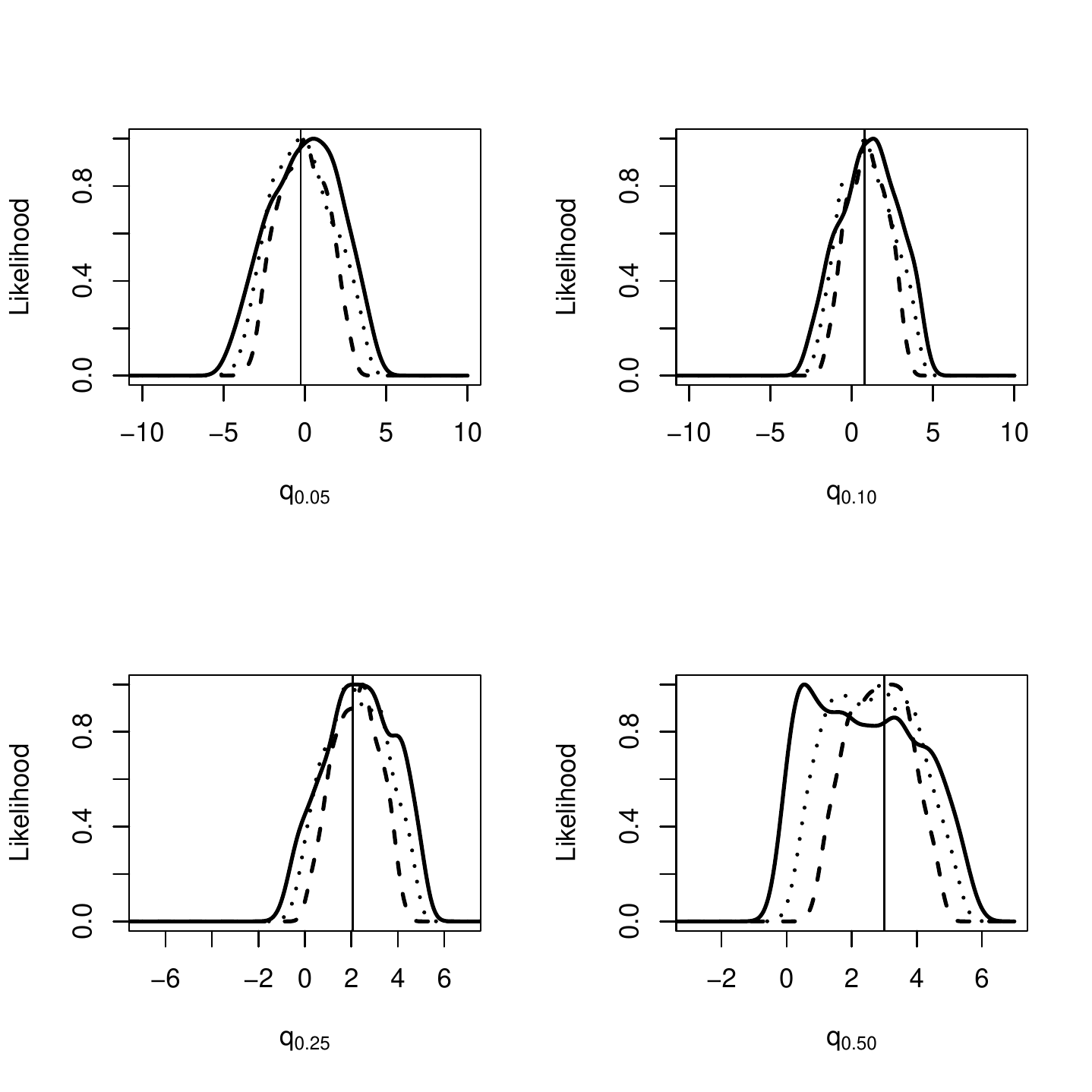}\label{fig:quantile - veros ABC}}

{\footnotesize Likelihood approximations of the quantiles of a $g$-and-$k$
distribution parameters for simulated data, obtained with an ABC algorithm
which simulates proposal values from the prior distributions 
($\mbox{U}\left(0,\,10\right)$
for each parameter): tolerance levels equal to $3$ (solid line),
$2$ (dashed line) and $0.5$ (dotted line).}
\end{figure}

\begin{figure}[h!]
\caption{\protect\includegraphics[width=9cm,height=8cm]
{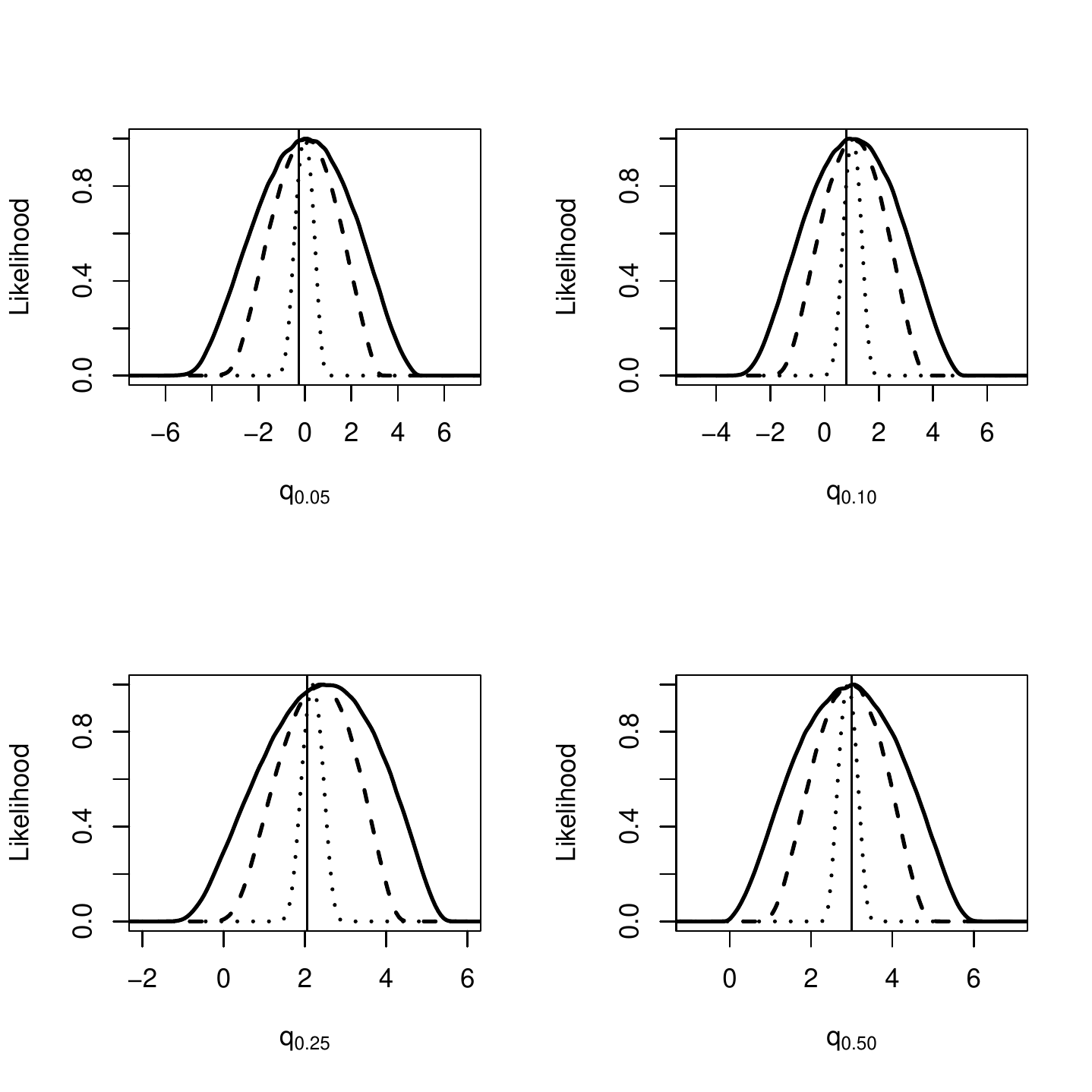}
\label{fig:quantile - ABCM veros}}

{\footnotesize Likelihood approximations of the quantiles of a $g$-and-$k$
distribution parameters for simulated data, obtained with an ABC-MCMC
Algorithm \ref{ABC-MCMC Mengersen} with Gaussian kernel: tolerance
levels equal to $3$ (solid line), $2$ (dashed line) and $0.5$ (dotted
line).}
\end{figure}

The results are shown in Figure \ref{fig:quantile - veros ABC}, \ref{fig:quantile - ABCR
veros}
and \ref{fig:quantile - ABCM veros}. The performance is in general very good:
the approximations are always concentrated around the true values. 

The ABC algorithm with simulations from the prior distribution has
some apparent problems of multi-modality, which are however absent using Algorithm 
\ref{ABC-MCMC Robert}.
However, in this case, the obtained approximations are not very smooth, and
they show more irregularities as the tolerance level decreases: as
we have already remarked, a too low threshold leads to very low acceptance
rates and this means that the chains do not move too much. 

In this example, Algorithm \ref{ABC-MCMC Mengersen} has the best overall
performance: the approximations are smooth and all concentrated around
the true quantile values. As the tolerance level decreases, the likelihood
approximations are more concentrated; obviously the computational
time gets larger.

The acceptance rates of these algorithms are in general very low:
\begin{itemize}
\item the basic ABC algorithm has an acceptance rate of $0.138\%$ when
the threshold is equal to $3$, and it goes down to $0.041\%$ and $0.007\%$
with tolerance levels of $2$ and $0.5$ respectively;
\item the ABC-MCMC Algorithm \ref{ABC-MCMC Robert} needs, respectively, $187$, $1487$,
about $500K$ and more than $3$ millions of simulations for the initialization step for the different
tolerance levels $3$, $2$, $0.5$ and $0.25$.The acceptance rates
of the proposed values is also very low: 18.41\%, 9.90\%, 0.47\% and
0.046\% respectively; it is clear that the acceptance rates relative
to the smaller thresholds cannot lead to smooth approximations;
\item the ABC-MCMC Algorithm \ref{ABC-MCMC Mengersen} needs $1104$, $4383$
and about $400K$ simulations for the initialization step for tolerance
levels 3, 2 and 0.5 respectively; in this case every accepted value
is a ``new'' value, and this solves
the problems in Algorithm \ref{ABC-MCMC Robert}. 
\end{itemize}
In conclusion, ABC-MCMC seems to perform better, although the versions
we have implemented present some cons: the algorithm in \citet{mar:12} is faster 
but it must be calibrated in terms of the tolerance
level, which has to be low in order to achieve good approximations, and the
MCMC acceptance rate, which has to be sufficiently high in order to allow the
chains to move. 

\begin{figure}[h!]
\caption{\protect\includegraphics[width=9cm,height=8cm]{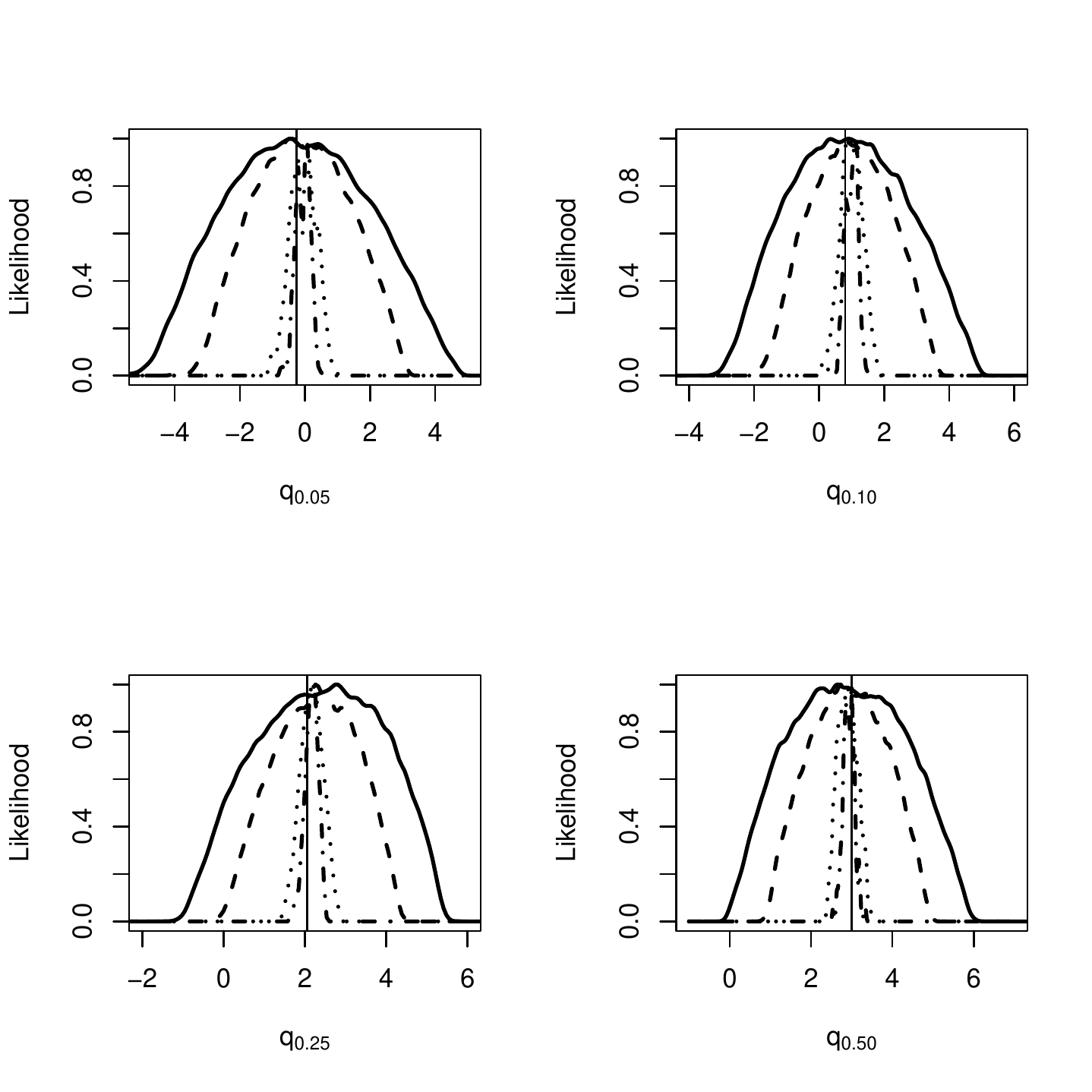}\label{fig:quantile - ABCR veros}}

{\footnotesize Likelihood approximations of the quantiles of a $g$-and-$k$ distribution
parameters for simulated data, obtained with an ABC-MCMC Algorithm
\ref{ABC-MCMC Robert} with Gaussian kernel: tolerance levels equal
to $3$ (solid line), $2$ (dashed line) and $0.5$ (dotted line)
and $0.25$ (dotdashed line). }
\end{figure}

\vspace{.21cm}

\noindent
{\bf Example 4.4}\,[Semiparametric regression]. Consider the following model
\begin{equation}
\bY=\bX \bbeta+ \gamma \left( \bz \right) + \varepsilon, 
\end{equation}
where $\bY=(Y_1, Y_2,...,Y_n)^\prime$ is  a vector  of $n$ real-valued variables, 
and $\bx=(x_1,x_2,  \dots, ,x_n)^\prime$,
$\gamma(\bz)=(\gamma(z_1), \gamma(z_2), \dots, \gamma(z_n))^\prime$ and $\bz=( z_1,z_2,\dots , z_n)^\prime$ are observed constants 
respectively taking values in $\mathbb{R}^p$ and $\mathcal{Z}$, 
$\varepsilon$   is the usual random component  that we assume having
multivariate normal distribution with mean $\bf 0$ and covariance matrix $\bf\Omega_\phi$ 
which depends  on some  parameters $\phi$,  $\bbeta$ is a vector of unknown parameters 
taking values in $\mathbb{R}^p$ and $\gamma: \mathcal{Z} \rightarrow \mathbb{R}$ is an 
unknown function. 

If the analysis is focused on $\bbeta$ or $\bf\Omega_\phi$, $\gamma$ may be considered a 
nuisance 
parameter and a method to remove it from the analysis is needed. In particular, if a 
weight function for $\gamma$ based on a zero-mean Gaussian stochastic process with 
covariance function $K_\lambda \left( \cdot,\cdot \right)$ with parameter $\lambda$ is 
used, the vector $\left( \gamma\left( z_1 \right),...,\gamma\left( z_n \right) \right)$ 
has a multivariate Normal distribution with mean $\bf0$ and covariance $\bf\Sigma_\lambda$ 
and the integrated likelihood function of $\bbeta$ is 

\begin{equation}
\left| \bf\Omega_\phi + \bf\Sigma_\lambda \right|^{ -\frac{1}{2}} \exp 
\left\{-\frac{1}{2} 
\left( \bY-\bX\beta \right)^\prime \left(\bf\Omega_\phi+\Sigma_\lambda \right)^{-1} \left( \bY-\bX\beta \right) \right\}
\end{equation}

\noindent where $\Sigma_\lambda$ is the $n\times n$ matrix with $K_\lambda 
\left( z_i, z_j\right)$ in the $\left(i,j \right)$ element.
This form may be obtained because of the assumption on the Normal distribution of the
 errors and the use of a Gaussian process weight function for $\gamma$; more general cases 
are not so straightforward to handle outside the Normal set-up.

In \citet{sev:13} the Authors  show that,  for a given choice of $K_\lambda \left(\cdot, 
\cdot\right)$, when the dispersion parameter, say $\boldsymbol{\eta}=\left(\phi,\lambda\right)$ is known, $\beta$ can be 
estimated by 
the generalized least-squares estimator: 
$\hat \beta=X^T \left(X^T V^{-1} X \right)^{-1} 
X^T V^{-1} Y$ where $V=\bf\Omega_\phi+\bf\Sigma_\lambda$; if the dispersion parameter 
is unknown, $\beta$ can be estimated as a function of an estimator of $\boldsymbol{\eta}$, $\hat{\beta} 
\left( \hat{\boldsymbol{\eta}}\right)$.

The method has been used with data from a survey of the fauna on the sea bed lying 
between the Queensland coast and the Great Barrier Reef; the response variable analysed is 
a score, on a log weight scale, which combines information across the captured 
species; this score value is considered dependent on the latitude $\bx$ in a linear way 
and on the longitude $\bz$ in an unknown way; see  \citet{ab:97}  for more details.
The model is
\begin{equation}
Y_j=\beta_0+x_j\beta_1 +\gamma\left(z_j\right)+\varepsilon_j,\quad j=1,...,n
\end{equation}

\noindent where $\varepsilon_1,...,\varepsilon_j$ are independent normal errors with mean 
$0$ 
and constant variance $\sigma^2_\varepsilon$. Using the integrated likelihood approach, a 
Gaussian covariance function 

\begin{equation}
K\left(z,\tilde{z} \right)=\tau^2 \exp \left(-\frac{1}{2} \frac{\left| z-\tilde{z}\right|^2}{\alpha}\right)
\end{equation}

\noindent and a restricted maximum likelihood estimate (REML, \citet{har:77}) for the nuisance parameters, the estimates of $\beta_1$  
is 1.020, with a standard error of 0.356 (see \citet{sev:13}).

We have used our ABC approximation in order to find an integrated likelihood for $\bbeta$. It is then necessary to define proper prior distributions for all the parameters of the model, 
i.e. 
$\boldsymbol{\beta}$, $\sigma^2_\varepsilon$ and the parameter of the covariance function 
of 
the Gaussian process, $\alpha$ and $\tau^2$. 

For $\boldsymbol{\beta}$ a g-prior has been chosen such that
$\boldsymbol{\beta} \sim\mathrm{N}_2 \left(\textbf{0},\, 
g \sigma^2_\varepsilon\left( \mathbf{X^T X} \right)^{-1} \right)$, where $g\sim \mbox{U}
\left(0, 2n \right)$ and 
$\sigma^2_\varepsilon \sim IG \left(a,b \right)$ with \textit{a}, and \textit{b} 
suitably small (as an approximation of the Jeffreys' prior).
A Gaussian process with squared exponential covariance function has been used as prior 
process for the 
function $\gamma\left(\cdot \right)$. The hyper-parameters of the Gaussian process have 
the following prior distributions: $\tau^2\sim IG\left(a,b \right)$, with $a=b=0.01$
and $\alpha\sim IG \left( 2,\nu \right)$ with 
$\nu=\rho_0/\left(-2 \log(0.05) \right)$ and $\rho_0=\max_{i,j=1...n}|z_i-z_j|$; see 
\citet{schgel:03} and \citet{BCG:04}  for more details. 

The choice of the summary statistics is not straightforward, because it is necessary to 
find 
statistics that take into account both the parametric and the nonparametric parts of the 
model, nevertheless sufficiency is not guaranteed. A function of \textit{z} has been 
considered and the maximum likelihood estimates of the coefficients of the new model have 
been used as summary statistics. In particular, two choices of function has been 
considered: $g\left(z_j\right)=z_j$ and $h\left(z_j\right)=z^2_j$ for $j=1,...,n$. An 
analysis of the maximum likelihood estimates has shown that the estimate of the constant 
$\beta_0$ is particularly unstable, therefore only the estimates for the predictor 
variables' coefficients contribute to the approximation as summary statistics.

In the MCMC step, normal transitional kernels have been used for all the parameters of 
the model, centred at the values accepted on the previous step and with small variance. 

\begin{figure}
\caption{\protect\includegraphics[width=9cm,height=8cm]
{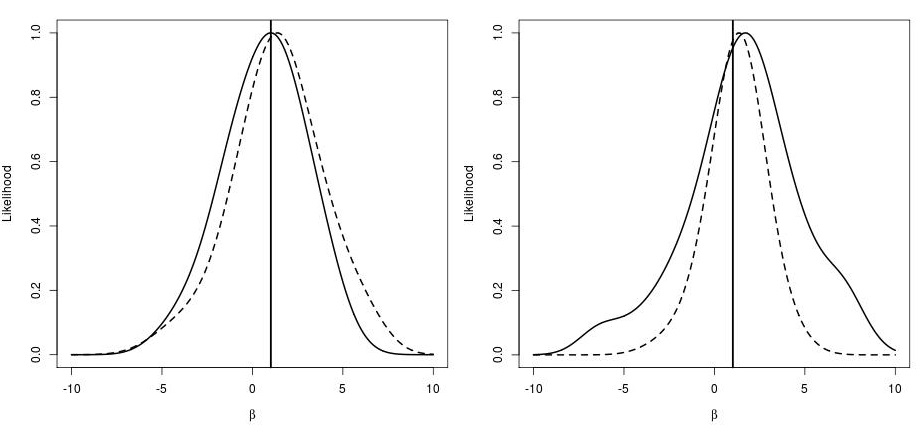}
\label{fig:semipar-ABC}}

{\footnotesize ABC approximation of the integrated likelihood of  $\beta_1$ in the 
semiparametric model. The approximations are obtained by using a Uniform prior (left) and 
an approximation of the Jeffreys' prior for $\sigma^2_\varepsilon$. Two different choices 
of summary statistics are compared: the maximum likelihood estimates of the model, with 
linear (solid lines) and quadratic (dashed lines) approximations of 
$\gamma\left(\cdot\right)$ and tolerance level of 0.5 (left) and 1.25 (right).}

\end{figure}

The results are shown in Figure \ref{fig:semipar-ABC}: the ABC approximation with $10^6$ 
simulations are concentrated around the estimates obtained by maximizing the integrated 
likelihood of the model. In this case, the ABC approach may be seen as a way to 
properly account for the uncertainty on the nuisance parameters that is not considered 
when  REML estimates are used.
Figure \ref{fig:semipar-ABC} compares different choices of summary statistics and prior distributions for 
the variance $\sigma^2_\varepsilon$: on the left a $ \mbox{U}\left(0,10 \right)$ is used 
and on the right a proper approximation of the Jeffreys' prior is used ($ 
\mbox{Ga}\left(a,b \right)$ with $a,b$ small). All the approximations are 
smooth and concentrated around the maximum likelihood estimate. Moreover, Figure \ref{fig:semipar-ABC} 
shows that using the summary statistics based on a quadratic approximation of $\gamma 
\left(\cdot \right)$ leads to better results, because they are all smooth. On the other 
hand the approximations obtained by considering a linear model with respect to $z$ present 
slight multimodality problems.

The number of simulations for the initialization step depends on the choice of the 
tolerance level: the approximation of the likelihood of $\beta_1$ by using a Uniform prior 
for $\sigma^2_\varepsilon$ needs 368, 2053 and 10945 simulations to accept the first 
value for tolerance levels of 1, 0.5 and 0.25 respectively; the approximation with Gamma 
prior with small parameters for $\sigma^2_\varepsilon$ needs 40, 34 and 81 simulations to 
accept the first value. These results refer to the summary statistics obtained with the 
quadratic approximation of $\gamma\left(\cdot\right)$, the other choice of summary 
statistics considered has shown similar values.

The acceptance rates of ABC-MCMC algorithm are in general low, in particular with the 
lowest tolerance levels;
 they are around $25\%$ for the highest thresholds considered.

\section{Discussion}
\label{5}

We have explored the use of ABC methodology, a relatively new computational tools for 
Bayesian inference in complex models, in a rather classical inferential problem, namely 
the elimination of nuisance parameters. We stress the fact that there are many situations 
where it is practically impossible to obtain a likelihood function for the parameter of 
interest in a closed form: in those cases the proposed method can be a competitive 
alternative to numerical methods.

As a technical aside one should note that, in many situations the prior $\pi(\bpsi)$ 
might be available in a closed form, so the kernel approximation of the prior is not 
necessary, and the accuracy of our method is even better. However, we have preferred to 
present the method in its generality.

 Another issue related to this last point is the approximation of the marginal posterior density of some components of the parameter. Also in this case, the problem is made simpler by the fact that no approximation is needed for the prior and standard asymptotic arguments for kernel estimators hold for the approximation obtained from the posterior sample.
The main drawback of the present approach is that it requires the use of proper prior densities.
This can be a problem, especially when the nuisance parameter is high-dimensional and the 
elicitation process would be difficult. A practical solution in these case is to adopt 
proper priors which approximate the appropriate improper noninformative prior for that 
model.


\end{document}